\documentclass{article}

\usepackage{arxiv}
\usepackage{graphicx}
\usepackage{float}
\usepackage[version=4]{mhchem}

\usepackage[hidelinks]{hyperref}
\title{A blockchain-based user-centric emission monitoring and trading system for multi-modal mobility}

\author{%
    Johannes~Eckert\\
    Machine Learning for Smart Mobility\\
    DTU Management, Transport Division\\
    Technical University of Denmark\\
    Anker Engelunds Vej 1, 2800 K. Lyngby, Denmark\\
   \texttt{johannes.eckert@ymail.com}\\
    \And
    David L\'opez\\
    GITraL, Universidad Nacional Autónoma de México\\
    Mexico City, Mexico City, Mexico, 04510\\
    LiTrans, Ryerson University\\
    Toronto, Ontario, Canada, M5B 2K3\\
    \texttt{dlopezfl@iingen.unam.mx}\\
    \And
    Carlos~Lima Azevedo\\
    Machine Learning for Smart Mobility\\
    DTU Management, Transport Division\\ 
    Technical University of Denmark
    \\Anker Engelunds Vej 1, 2800 K. Lyngby, Denmark\\
    \texttt{climaz@dtu.dk}\\
    \And
    Bilal~Farooq\thanks{corresponding author}\\
    Laboratory of Innovations in Transportation (LiTrans)\\
    Ryerson University\\
    Toronto, Ontario, Canada, M5B 2K3\\
    \texttt{bilal.farooq@ryerson.ca}\\
}

\begin{document}
\maketitle

\begin{abstract}

Since the transport sector accounts for one of the highest shares of greenhouse gases (GHG) emissions, several existing proposals state the idea to control the by the transportation sector caused GHG emissions through an \emph{Emission Trading Systems} (ETS). However, most existing approaches integrate GHG emissions through the fuel consumption and car registration, limiting the tracing of emissions in more complex modes e.g. shared vehicles, shared rides and even public transportation. This paper presents a new design of a user-centric ETS and its implementation as a \emph{carbon Blockchain framework for Smart Mobility Data-market} (cBSMD). The cBSMD allows for the seamless transactions of token-equivalent GHG emissions when realizing a trip, or an emission trading action as well as the transaction of individual, service or system-wide emission performance data. We demonstrate an instance of the cBSMD implementation for the transactions of an ETS where all travellers receive a certain amount of emission credits in the form of tokens, linked to the GHG price and a total emission cap. Travellers use their tokens each time they emit GHG when travelling in a multi-modal network, purchase tokens for a given trip when they have an insufficient token amount or sell when having a surplus of tokens due to a lower amount of emitted GHG. This instance of cBSMD is then applied to a case-study of 24hours of mobility of $3,186$ travellers from Oakville, Ontario, Canada, where we showcase different cBSMD transactions and analyze token usage and emission performance.
\end{abstract}


\section{Introduction}
%Greenhouse gases (GHG) are necessary for sustaining life on Earth as they reflect most of the infrared radiation from the atmosphere back to the Earth. Without these gases the Earth's surface temperature would be around -18°C instead of the current average temperature of 15°C) \cite{Jain1993}.
%However, the increasing global population and the consequently higher demand in nourishment, energy and transportation possibilities has resulted in a greater GHG emission and therefore in a stronger reflection of infrared radiation \cite{Masson2018}.

Many governments have implemented \emph{Emission Trading Systems} (ETS) in order to meet the \emph{Paris Agreement} and \emph{Kyoto Protocol} which strive for a constant reduction of the global Greenhouse Gas (GHG) emissions. ETS promises a direct GHG quantity control mechanism where reduced emission allowances are allocated to emitting as well as possibly non-emitting stakeholders and traded through different possible schemes \cite{EC2008}. Therefore, the \emph{Kyoto Protocol} has a strong potential to affect the carbon footprint of individuals, but it forces reductions in the production process rather than the consumption process \cite{Pan2019}. 

The causality between number of cars and GHG emissions has a main impact on the global warming. During the last decades the number of motor-driven vehicles has continuously increased and more than 1 billion vehicles are on the global streets nowadays \cite{Stacy2019}. The result is that the transport sector accounts in many countries one of the highest shares of GHG emissions, for example, with $28.9\%$ \cite{EPA2017} and $24.3\%$ \cite{Eurostat16} in the U.S and the E.U respectively. While some countries have tried to include maritime and air transport in its proposed ETS, the GHG share by the road sector is tremendous compared to these sectors \cite{ipcc2017}. Furthermore, in urban areas the pollution by the road sector is the highest due to its concentrated population ($75\%$ of all people in the industrialized and $40\%$ in the developing countries) and the traffic situation propels people to a non-efficient driving behavior, which leads to higher GHG emissions \cite{ipcc2017}. Thus, some countries and cities have started to announce actions against fossil fuel vehicles in order to reduce and avoid GHG emissions from the road transportation sector \cite{World2010}. Those actions, for instance, sanctions against older vehicles or the implementation of a carbon tax, whereby users have to pay an additional share per liter gasoline \cite{DFC2018}, are supporting the declarations aiming at fighting the global warming, as the \emph{Kyoto Protocol} and the \emph{Paris Agreement}. 

An ETS relies on a supply and demand law due to a \emph{cap and trade} mechanism for GHG emissions \cite{EC2008}. Quantities of future GHG emissions are distributed across individuals and each participant in the system receives a certain amount of emission allowances. These allowances can be traded as far as a user changes his environmentalism and produces less GHG emissions than his allowances permit. There are several different ETS operating around the world. For instance, the world largest ETS is in the European Union (EU) \cite{EC2008}. The EU issued emission allowances for more than 11,000 heavy energy-consuming organizations in 2005. Thus, all low emitters, who reduced their GHG emission by implementing new technologies or changed procedures, can trade their surplus on allowances to other companies which have run out of GHG allowances for this year. The annual, linear reduction of allowances incentives the participants to actively reduce the GHG emissions. 

Furthermore, ETS arouse more and more attention from the transportation sector. In combination with the three different streams of the transportation sector - the upstream, midstream, and downstream \cite{Mock2014, Iankov2008} - new approaches have been made. The upstream contains the fuel producers, who are responsible for the fuel's emission factor, which affects the amount of GHG emissions due to combustion. According to \cite{Grayling2006} the most effective and political accepted approach would be to include the fuel producers (upstream) in carbon trading mechanisms as the EU ETS. The midstream is accountable for the vehicle fuel economy. Thus, \cite{Winkelman2000} state that the vehicle manufactures (midstream) should be responsible for the GHG emissions which are produced through the sold vehicle. \cite{Dobes1998}, however, describes a permit scheme in which the vehicle users (downstream) are responsible for the emissions due to the use of combustion engines. %As a solution, every user gets permits to refuel his car \cite{Dobes1998}.

The blockchain technology is often proposed in order to increase the security, end-user control, and transparency in ETS. Blockchain belongs to the superior Distributed Ledger Technology (DLT), which is characteristic to store, distribute and facilitate the exchange of value between users without a central authority \cite{Beck2017}. Furthermore, the blockchain is formed by linked blocks which are in a chronological order and contain transactions \cite{Farooq2019}. Those linkups between the blocks are permanent and cannot be tampered after the transaction was made. This mechanism, in connection to the consensus algorithm, which allows for the distributed flow of information, secures the transactions \cite{Dapp2017}. 
%Blockchain gained in almost every industry sector a lot of attention during the last years. 
For the transportation sector, especially the new area of electrified, connected and autonomous vehicles indicates huge possibilities for blockchain-based implementations. Renewable energy, for instance, must be used when generated, the unpredictable generation of energy (e.g. through windmills) can end up in a disconnection of the grid \cite{Pop2018}. Blockchain can help to develop a transparent way to balance the energy level at the grid level without a third party \cite{Pop2018, Nguyen2016}. Consequently, it can be rewarded when the user charges the car while the grid level is high and the other way around, respectively. The energy information is verified by other peers in the system, the real amount of generated energy, the origin as well as the current price is transparent. Blockchain can be used whenever the sharing of data is needed where the transaction record is stored on a persistent distributed ledger. As a result of this automatic record keeping, the technology entitles individuals to do more with their own resources \cite{Sharma2017}. For example, parking applications can involve private parking lots and garages to share with other participants in the system \cite{Parksen2018}.
%According to \cite{Yuan2016} the blockchain technology has the power of revolutionizing the current \emph{Intelligent Transportation Systems} (ITS) and their mostly centralized \emph{Internet of Things} (IoT) applications. Therefore, the gathered mobility data can be stored instead in a centralized data bank in a decentralized ledger. Consequently, physical devices could be transferred with embedded chips to mining platforms. Those tamper-proofed records in the ledger could result in, for instance, lifetime data logs of cars.
Nevertheless, heretofore there are just theoretical approaches of blockchain implementations in carbon trading schemes \cite{Liss2018, Pan2019, Wenxiang2019}. According to \cite{Liss2018} a blockchain implementation to the EU ETS would be beneficial. Several hacker attacks and unclear transactions results in a damage around 5 billion euros for the EU \cite{Auditors2015}. The blockchain could eliminate the centralized system which does not monitor financial information nor prices of the transferred allowances nowadays. However, the application of blockchain in personal carbon trading systems will accelerate the people's involvement in carbon trading and sensitize the society to the personal caused carbon footprint.

Recently, \cite{Wenxiang2019} proposed a conceptual design of a blockchain-based ETS for the road transport sector integrating the upstream, midstream and downstream stakeholder. Similarly to most approaches, \cite{Wenxiang2019} proposed the control of carbon footprints at the downstream level through fuel transactions. In addition, vehicle ownership and fuel production has also been considered in their ETS concept. However, such design entails that the GHG emissions are not assigned directly to individual's trips, limiting the tracing of emissions in shared vehicles, shared rides and even public transportation that ultimately (and increasingly) contribute to the sector's total emissions.

Moreover, existing proposals are not implying the time factor, thus the amount of GHG emissions cannot be assigned to individual trips. Although time-stamped data and real-time recordings can be used for a dynamic GHG management and trading system, respectively. For these purposes, the blockchain technology has the potential to maintain a transparent and secure monitoring and trading system and therefore to incentivize individuals to reduce their personal carbon footprint. 

This paper presents the development of a blockchain-based ETS in a user-centric and multi-modal perspective. It extends existing state-of-the-art conceptual frameworks (namely Li et al. \cite{Wenxiang2019}) with trip- and user-specific tokenized emission credit transactions; it presents its architecture and implementation under an open-source \emph{carbon Blockchain framework for Smart Mobility Data-market}, cBSMD\footnote{https://github.com/LITrans/cBSMD}; and showcases its application in a case-study. The structure of the remaining paper follows these three key contributions with dedicated sections.

\section{cBSMD Framework}
The conceptual framework was built on the basis of the \emph{Multi-layered blockchain framework for smart mobility data-markets} (BSMD) by \cite{Farooq2019}. The approach tackles the fact that most of the mobility data are stored in centralized servers, what makes it easy for hacker attacks as well as for undesirable shares with third parties \cite{Farooq2019}. Therefore, the BSMD gives each user the power to set up his transaction rules. Consequently, the encrypted data will be just transferred when both, the owner of the data and the receiver of the data, agree on the terms of the smart contract. To sum up, the BSMD helps to secure the data ownership, ensure the transparency and auditability of transfers, and intensify the access control to personal gathered mobility data. 

The original BSMD is structured in six layers. 1) The \emph{identification layer}, 2) \emph{privacy layer}, 3) \emph{contract layer}, 4) \emph{communication layer}, 5) \emph{incentive layer} and the 6) \emph{consensus layer} \cite{Farooq2019}.

In the following sections we present the architecture of the proposed cBSMD along with the changes to the original layer implementation in order to align them with the GHG emission trading here at stake. We introduce a carbon token that is composed of the \ce{CO2e} (equivalent) of all types of GHG emissions. 
%The considered GHG emissions are composed of Water vapor (\ce{H2O}), Carbon dioxide (\ce{CO2}), Methane (\ce{CH4}), Nitrous oxide (\ce{N2O}), Ozone (\ce{O3}), Chlorofluorocarbons (CFCs), Hydrofluorocarbons (HCFCs and HFCs), Sulfur hexafluoride (\ce{SF6}). Therefore, all GHG emissions are converted to Carbon dioxide equivalent (\ce{CO2e}) to calculate with a consistent factor and to consider the impact of the different gases. As a consequence, a carbon token is composed of the \ce{CO2e}.

\subsection{Architecture}

The first step is to define the level of permission to determine the involvement of the participants. \emph{Open} systems consume a high amount of energy due to the great involvement of participates in the consensus mechanism \cite{Vranken2017}. Aligning with the fundamental goal to control GHG emissions, only \emph{closed} systems should be considered. 
Moreover, since the scheme is in particular for the public, participants should not have any restrictions in terms of access the ledger or do transfers within the system. This leads to the same type of permission level as the BSMD system contains: \emph{public closed}.

The users, anyone who takes a trip, is entitled to access the system and participate in the GHG trading mechanism. However, each user has to participate in the exchange of GHG emissions to ensure an equitable ecosystem. The cBSMD not only considers the emitted GHG per car, but also the GHG emissions per user. This entails that the user is linked to his current mode while traveling (Figure \ref{fig:cBSMD_framework}). Hence, the number of passengers is crucial to determinate the accurate GHG footprint of the individual user. Therefore, the vehicle takes the total amount of emitted GHG and divides the number by the current quantity of passengers. As a consequence, the mode is entitled to record the emissions on behalf of the user to his \emph{identification}. However, bus has to be considered as an exception. Even though the operation of buses, especially off-peak times, are inefficient. A charge by number of passengers would be disproportionate for the individual user. Therefore, the authors suggest a charge per seat system for the trips made by bus. This charge per seat system could be also implemented for other shared vehicles. As a result, the carbon tokens for remainder seats could be paid by the operator. Thus, a higher efficiency of operators would be inducted.

\begin{figure}[!h]
    \centering
    \includegraphics[width=0.4\textwidth]{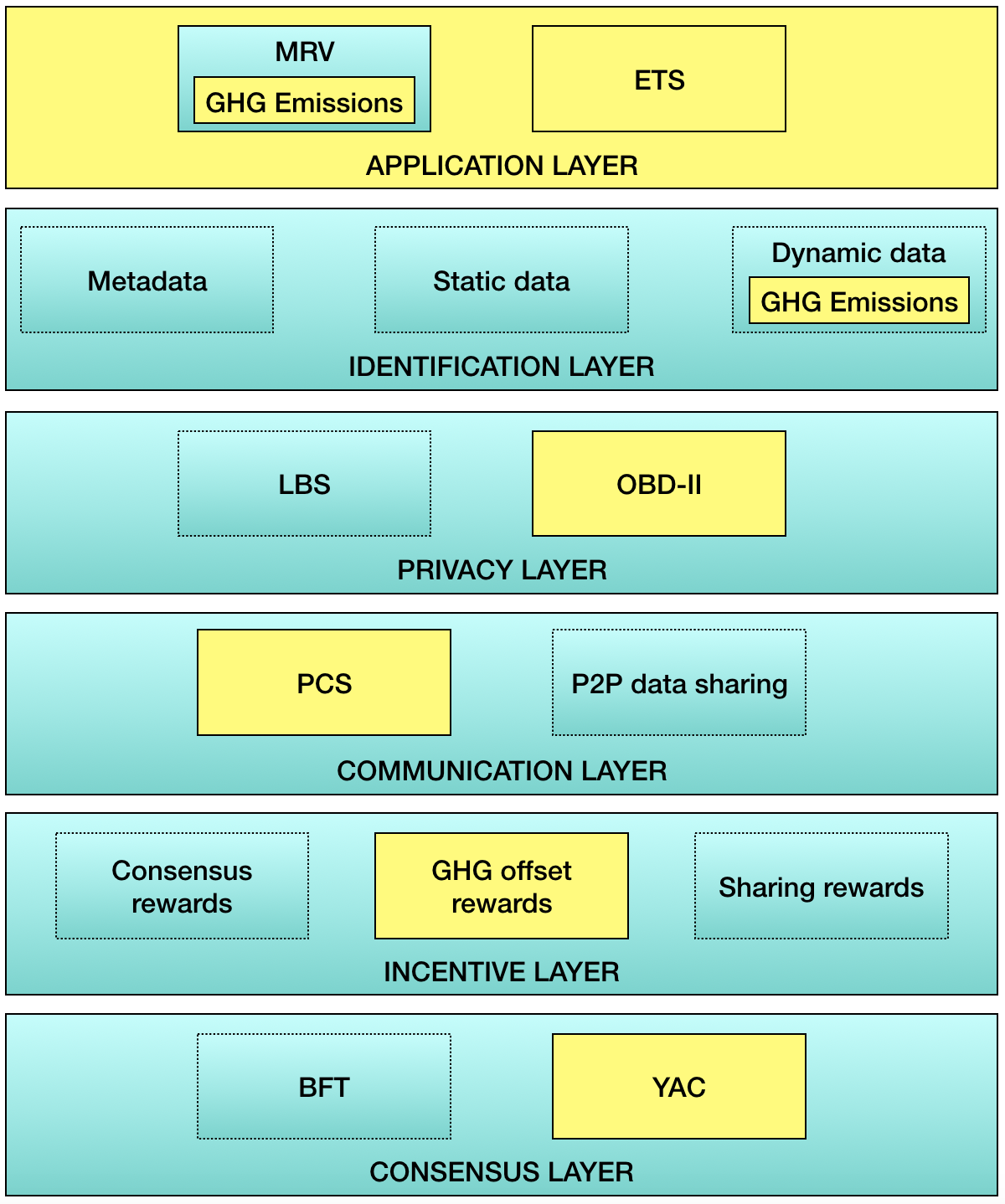}
    \caption{Structure of the cBSMD (yellow represents the contributions of this work in the BSMD)}
    \label{fig:layers}
\end{figure}

As a norm in blockchain systems \cite{Yuan2016}, the cBSMD is described as the BSMD in layers (Figure~\ref{fig:layers}). As mentioned before the cBSMD is an expansion of the BSMD which uses the principles of the blockchain system and assigns personal GHG emissions to the users. The changes and additions towards the original BSMD are highlighted in Figure~\ref{fig:layers}. 
A new \emph{application layer} was added to illustrate the different features of the system and for which purpose they can be used. The \emph{identification layer} is composed by the sole owner possessed information. The \emph{privacy layer} describes which type of data would be required for the cBSMD and how the privacy of each user can be ensured. The \emph{communication layer} is composed of the communication processes between nodes in the system. The \emph{incentive layer} demonstrates the way how people can be incentivized to participate in the system passively and actively respectively. Lastly, the \emph{consensus layer} encloses the used algorithms which allows active nodes, by given circumstances, to write transactions in the ledger. The contract layer of the BSMD was discarded due to the fact that at the current status no smart contracts and brokers are needed for the exchange of GHG tokens. 

\subsection{Application Layer} \label{layers}

\begin{figure}
    \centering
    \includegraphics[width=1\textwidth]{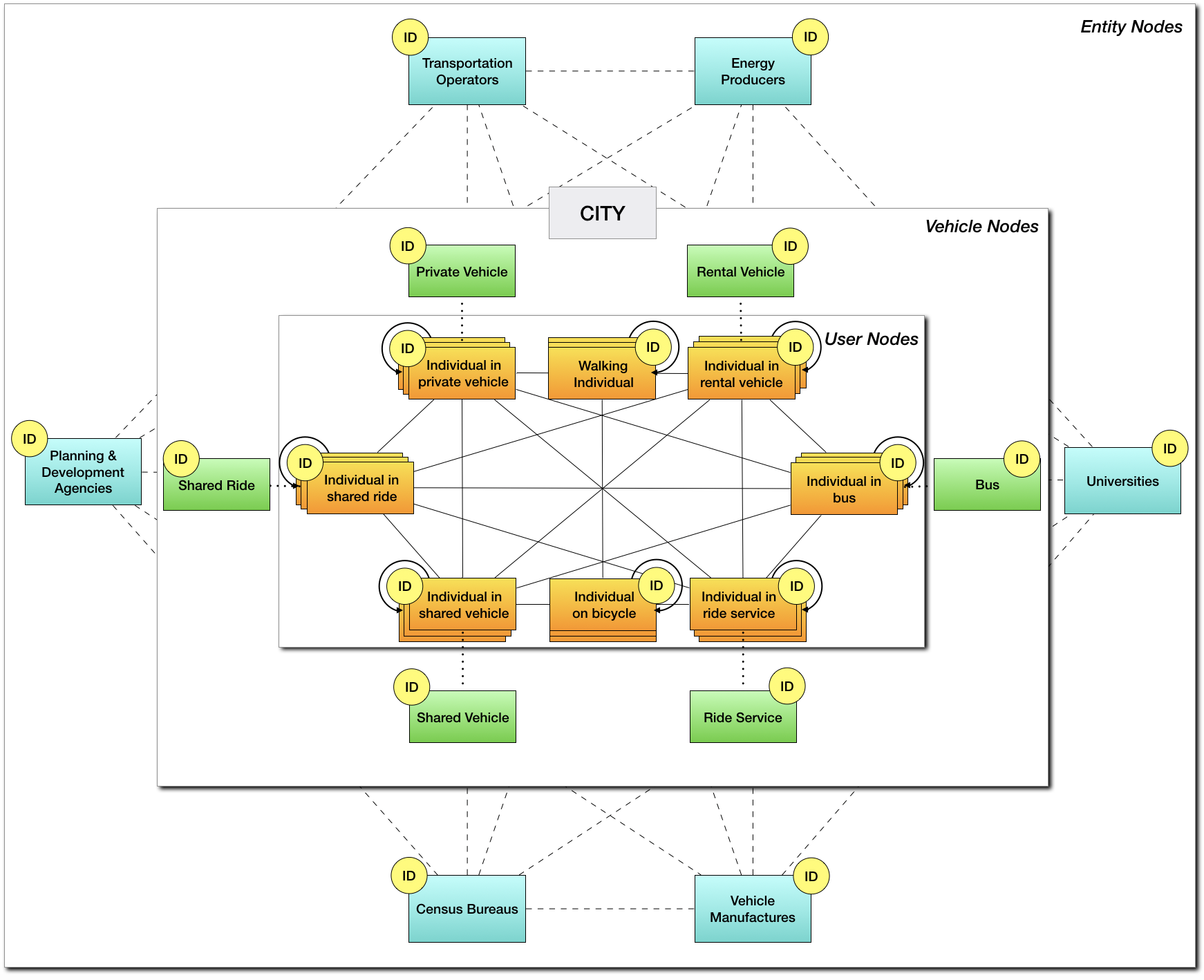}
    \includegraphics[width=1\textwidth]{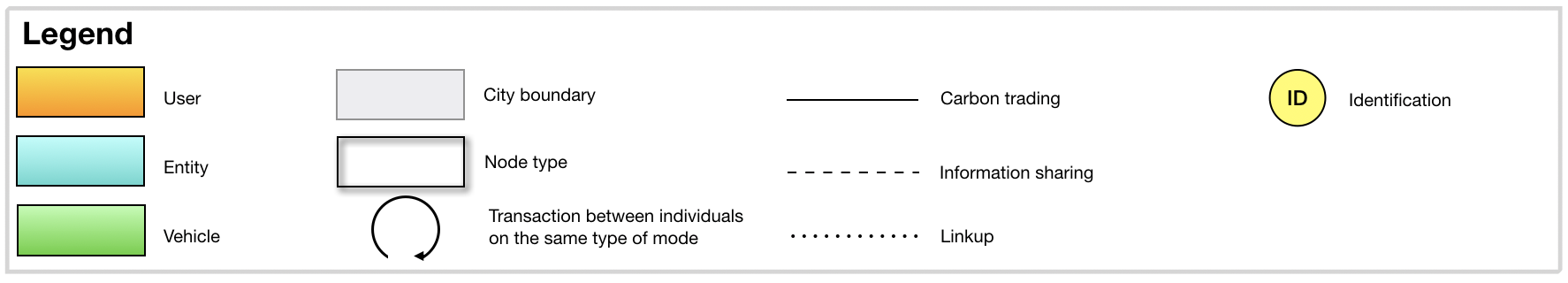}
    \caption{Conceptual framework of the cBSMD}
    \label{fig:cBSMD_framework}
\end{figure}

The application layer \cite{Yuan2016} was added to the structure of the existing BSMD to highlight the new proposed feature of the cBSMD, the ETS. Figure~\ref{fig:cBSMD_framework} illustrates the general framework of the cBSMD and contains the main functions of the system. At first, the transparent and accurate monitoring, reporting and verification (MRV) feature. Similar to the original BSMD scheme, the user can store his trip-data to his \emph{identification}. Thus, the sole owner of the data has always an overview about his trip-based information as, for example, travel time, travel distance and mode. In addition to those mobility related data the personal GHG footprint of the users was included. Furthermore, the node with the ownership of the information can allow other nodes to acquire insight to particular information written in the ledger. As a consequence, a user can share now in addition to his travel data his \ce{CO2e} footprint with, for example, a university for research purposes, a policy maker for spatial analysis or a trip-specific mobility operator (e.g. public transit, car share operators). However, not just users can share their GHG emissions, also the vehicle itself. %This entails an accurate MRV system for mobility operators and other entities. 

Secondly, the novelty, the user-centric emission trading function. In order to control the GHG emissions in a given boundary, an transaction process was implemented. The transactions of tokenized emissions allows the record of tokens used in each trip, and the trade of tokens between participants when consuming (needing) less (more) than the available token budget. 
Therefore, the node contains alongside each trip record, the respective token (or credit) information, the emitted GHG quantity and its equivalent in terms of current \ce{CO2e} price. Consequently, accurately measuring or estimating the amount of emitted GHG during a trip is crucial for the number of tokens which are needed to cover the trip. A detailed description of the implementation of the GHG trading mechanism can be seen in section \ref{Implementation}. 

\subsection{Identification Layer}
Nowadays, each person and each vehicle generate, while traveling, a certain amount of data. This data can be used by entities by way of example for research purposes, statistical evaluations or planning and operational processes. The initial approach of the BSMD distinguished therefore between metadata, static data and dynamic data which is stored in the \emph{identification} of each node. 
The metadata is always accessible for other nodes in the system. As a result, another node can see the type of node (user, vehicle, entity) and the blockchain address to get in touch with this node. However, the metadata does not disclose any personal data and ensure the sole ownership of sensible data. 
Secondly, the static data is composed of all the information which does not change often. Typical static data is for example the information about the name, gender, address. For entities or vehicles the static data consists, for instance, the company name or vehicle brand. Moreover, due to the presentation of a by a government authority issued document (e.g. birth certificate, passport) the authenticity of the node’s metadata is ensured and the static personal data (age, gender, address, etc.) verified. Finally, the dynamic data in the \emph{identification} describes the gathered trip related information. In this dataset the cBSMD added the information about the GHG emissions of the trip as well as the equivalent amount of tokenized carbon credits. Furthermore, the emission trading scheme leads to additional information about executed transactions in the dynamic dataset, which includes data and time of the transaction, the sender's and receiver's blockchain address, if applicable a description with the vehicle id, and the amount of transferred tokens. 
This leads to a total overview about statistical trip data, GHG emissions as well as the transactions which were required to offset the personal trip-based GHG footprint and describe the savings on emitted \ce{CO2e} respectively. 

\subsection{Privacy Layer}
%The privacy layer contains the required data sources and the level of privacy to this data. 
As the BSMD already declared, the Location Based Services (LBS) are implemented in many different applications to inform the user about the trip itself (e.g. travel distance) or to provide information about the environment (e.g. location of the next gas station). However, the BSMD implemented with \emph{geomask} and \emph{GeoInd} possibilities to set level of privacy to conceal or disclose the current location \cite{Farooq2019}. This hybrid privacy approach can also be implemented in the cBSMD.
However, for the cBSMD the LBS is not only important to share trip related information with other nodes. Additionally, the LBS ensures the validity of trips. Thus, users of the ETS have to be indicated by a start and end location. This concerns especially users who take an emission-free alternative for their trip, as walking or cycling. Also users in vehicles have to be assigned to a start and end location of the trip due to the possibility that the vehicle does not stop the ride when the user does (i.e. bus and shared ride trip). Hence, the cBSMD has to rely, for example, smartphones and other ITS technologies in order to ensure the current position of the user. 

%\hl{Usually ETS are set by a \emph{cap and trade} mechanism. In original industry-based ETS is this granted by the affiliation of companies to countries or greater areas as the EU, in which the participates can trade their allowances under the set cap. In order to implement a meaningful cap to a user-based ETS in the mobility sector the LBS could provide the required information. This entails, for instance, whenever a vehicle enters the city boundary the user enters simultaneously the cBSMD to share his emission data to entities, if confirmed in the user's \emph{identification}, and participate in the GHG trading mechanism.}
A typical approach for estimating vehicle emissions is to use driving-cycle-based models (e.g. MOBILE6, EMFAC and measure the tailpipe emissions for these standard cycles \cite{Frey2003}. With this data set, average emissions can be estimated for the vehicle's average speed. 
However, in order to consider the actual GHG emissions by the vehicle and to take into account driving behavior as well as the condition of crucial vehicle parts, the \emph{On-board diagnostics} (OBD) system could be used to implicit a accurate GHG monitoring and tradings system. Nowadays GHG emissions of vehicles are measured only via the OBD-II (latest and normalized version of OBD) system. Moreover, tailpipe measurements regarding the GHG value are prohibited since the Volkswagen emissions scandal in 2015. However, the accurate emission data, analyzed by the OBD-II system in the vehicle, can be transferred to the \emph{identification} of the vehicle. In this case a vehicle can share his GHG emissions without disclosing any sensible data as the GPS signal, or travel history. Furthermore, the vehicle can reduce the user's amount of carbon tokens based on the information of the OBD-II. Therefore, the total amount of GHG emissions is divided by all linked passengers to the vehicle. Consequently, the vehicle does not only store the GHG footprint in the vehicle's \emph{identification} it can also write the GHG amount, divided by the number of passengers, on behalf of the users to theirs \emph{identifications}.

Although the GHG emission information has to be shared with other nodes, the level of privacy is not affected negatively. The personal data can be secured and the GHG emission consumption is anonymized due to the use of public accessible \emph{metadata} instead of personal information.

\subsection{Communication Layer}
The cBSMD was built with \emph{Hyperledger Iroha}. Therefore, the \emph{Peer Communication Service} (PCS), an intermediary, is part of the cBSMD. The PCS establishes a communication channel between two nodes, whenever both nodes accept the requirements of sharing information. Thus, the PCS is responsible for the data transition from the entry point for users via the \emph{Multisignature Transaction} (MST) Processor through to the \emph{Ordering Gate}. The \emph{Torii}, the entry point for users, as well as the MST Processor use the open source remote procedure call\emph{gRPC} to connect the user with the system and other peers. In addition, the MST Processor exchanges the status of unsigned transactions among the users to reach the required quorum of signatures. Thus, each peer in the system receives messages from other peers’ trough a \emph{Gossip protocol}. Since a distributed ledger does not have a central registry, the \emph{Gossip protocol} establishes an epidemiologic communication service which distributes messages to nodes and their linked peers. Consequently, the PCS sends stateless but valid transactions to the ordering gate along other peers and can hide the complexity of interaction with the consensus implementation. Due to a fact that a single node has several communication channels, a hacker attack would have to intercept multiple communication connections before entering transferred data. In general, the PCS ensures a secure P2P data sharing system.

\subsection{Incentive Layer}
In order to incentive the participants, the incentive layer contains three different mechanisms. 
First of all, like the most blockchain systems, the involvement in the consensus mechanism to incentive nodes to participate in the consensus mechanism. Thus, a node is entitled to create a new block due to the given consensus mechanism, for instance, solve a difficult mathematical problem based on a cryptographic hash algorithm (proof-of-work) or collect votes (Byzantine Fault Tolerance) \cite{Farooq2019}. The node may receive for every successfully new created block the transaction’s fee as a reward.
Secondly, the reduction of GHG emissions. The users can be incentive by the system, in terms of reduce the personal GHG footprint. Whenever a user decides to emit less GHG, he receives rewards in terms of his surplus what he can share and his saving respectively. In the other way around, a user has to purchase more tokens as offset for his extra GHG emissions. This mechanism should incentive individuals to reduce the personal GHG footprint while receiving rewards or reducing expenses. 
Finally, the sharing rewards which is already implemented in the BSMD scheme. Those rewards are given to users who are willing to share trip related trip information. For instance, an individual can share his trip data to an entity for MRV purposed and get paid by the receiver.

\subsection{Consensus Layer} \label{Consensus Layer}
The consensus layer consists the cBSMD’s mechanism for the consensus process. This is in particular the \emph{Yet Another Consenus} (YAC) which uses the \emph{Byzantine Fault Tolerance} (BFT) consensus algorithm with an added modular architecture as well as a simple implementation structure \cite{Muratov2018}. YAC’s distinctive features are its scalability, low latency for transactions and high transaction throughput \cite{Muratov2018}.

The BFT is also the underlying consensus algorithm of the BSMD. The fault tolerance assumes that a small number of unreliable or potentially malicious nodes can be tolerated in order to ensure the validity of a transaction \cite{Castro1999}. If the stake of dissidents is no more than 1/3, the correctness of a block can be guaranteed. Nevertheless, the BFT algorithm is more suitable in trusted environments rather than in \emph{public open} systems due to the fact that transactions are verified by individuals and signed by known validator nodes \cite{Castro1999}. In order to reach consensus and create a valid transaction a certain amount of signatures have to be collected, minimum 2/3 of the participants in the consensus process.

\section{cBSMD Implementation} \label{Implementation}
\begin{figure}[!h]
    \centering
    \includegraphics[width=1\textwidth]{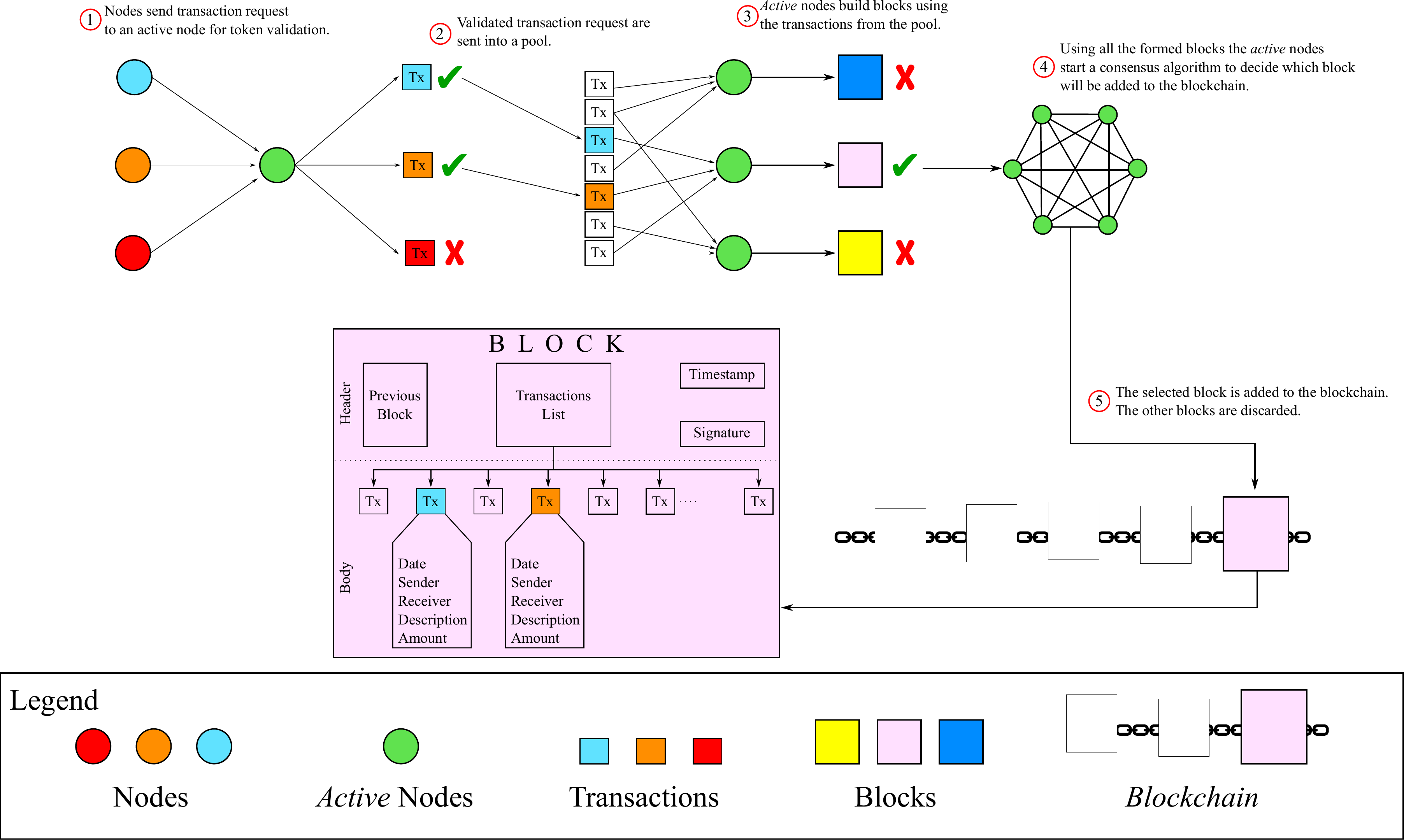}
    \caption{Life cycle of a transaction request in the cBSMD}
    \label{fig:block_in_ledger}
\end{figure}

To demonstrate how the cBSMD can be used as a trading system for GHG emissions we implemented the cBSMD nodes on \emph{Hyperledeger Iroha} (see Figure \ref{fig:block_in_ledger}). For this paper only token transactions are implemented since we are interested in how the tokens are used during the day. For the simulation we left out data transactions from users and cars to entities, which contain, for instance, the GHG emissions information or trip related data.

All users in the cBSMD are entitled to consume a certain amount of tokens. We first assume that at the beginning of the day all users have the same amount of tokens which are added to the users wallets. This assumption can easily be relaxed to allow for different allocation policies of the ETS. In the cBSDM ledger it is recorded the amount of tokens each user has at beginning of the day. Wallets in the cBSMD stores tokens and can only be altered (add or remove tokens) if a transaction request is sent and approved by the \emph{active} nodes. Every token transaction is always recorded in the cBSMD ledger. 

In the current status of the cBSMD we consider the payment of tokenized credits for the produced GHG after the trip is done. However, we will also take upfront costs into account for further steps. For now, when a user finishes a trip they must pay a proportional amount of tokenized credits for the produced GHG. For paying the trip the user sends a transaction request to the \emph{active} nodes where they inform that they are going to alter their wallet to pay the tokens they spent in the trip.
%The transactions request contains the following information:
%\begin{itemize}
%    \item \emph{Date}: date of the transaction
%    \item \emph{Sender}: public id of the node sending the tokens. Here it is the public id of the user who made the trip
%    \item \emph{Receiver}: public id of the node receiving the tokens. Here it is the public id of the agency who create the initial credits of the user. For the case to cBSMD are trading tokens this field will have the public Id of the user receiving the tokens
%    \item \emph{Description}: message attached to the transfer. This message may contain the GHG produced and summarized info of the trip.
%    \item \emph{Amount}: total number of tokens.
%\end{itemize}

The life cycle of a token transaction request is depicted in Figure~\ref{fig:block_in_ledger}. At first, the transaction requests is sent to one of the \emph{active} nodes in where it is verified; (1) if the transaction is well formed (i.e., is written in a language readable by \emph{active} nodes) and (2) if the user has enough tokenized credits for paying the trip. Since every token transaction is recorded in the ledger the \emph{active} nodes can easily verify if the paying node has enough tokens in his wallet. 

If the transaction is well formed and pass the token validation, it is sent into a pool where it waits to be groped into a block by an \emph{active} node. More than one \emph{active} node can build blocks using the transactions from the pool. Once a predetermined number of blocks are created, the \emph{active} nodes starts a consensus algorithm called YAC to decide which block will be added to the blockchain. The blocks that were not selected by the consensus are discarded. %All blocks added to the blockchain have the following information:

%\begin{itemize}
%    \item \emph{Previous block}: hash value that chains the block to the previous block
%    \item \emph{Transaction list}: array of transactions. Blocks contains more than one transaction
%    \item \emph{Timestamp}: current time when the block was created
%    \item \emph{Signatures}: signatures of \emph{active} nodes, which voted for the block during consensus round
%\end{itemize}

\section{Case study}
The trading system over the cBSMD is simulated on a physical network and with a blockchain composed of $3,186$ \emph{passive} nodes that represent users traveling in a day and 4 \emph{active} nodes in charge of hosting and running the cBSMD. An $i5$ $3.5$GHz$\times4$ and $8$GB of RAM do every transactions of the active nodes while 4 \emph{t2.medium} (Amazon cloud EC2 virtual machines with 2cores at 3.1GHz and 4GB of RAM), runs 1 \emph{active} node each.

To analyze sociodemographic related causalities in terms of the carbon token usage in the cBSMD, we take the survey data from the 2011 Transportation Tomorrow Survey (TTS) of Oakville, which is a suburban town of the Greater Toronto Area, Ontario, Canada. However, the case study does not reflect the total amount of transportation related GHG emissions from the area and the total carbon token consumption. In order to control GHG emissions through an ETS, a total cap is needed. For this exercise, the cap in the system is determined with the total amount of emitted GHG of trips which start and end in Oakville. This boundary is important for the subsequent implementation of a sufficient \emph{cap and trade} mechanism, whereby the cap of allocated tokenized credits can be continuously reduced. We can allocate an average carbon token budget to each traveling individual through the total amount of GHG emissions and the number of travellers. The case study shows how an allocation of carbon tokenized credits to individuals would affect the different groups of population. Based on the outcome of the case study, an evidence backed trading scheme for an individual GHG market can be applied. Note that for such demonstration purposes, only the real trips of the survey were considered without expansion factor to the population.

The scope of this study is to first analyze the trips by the users, calculate the total amount of carbon tokenized credits based on the emitted GHG emissions and allocate each user with the average number of carbon credits. After that we can simulate the need of tokens by user. As a consequence, we can dissect, based on sociodemographic and mobility-related data, which users have a higher GHG footprint than the average and runs out of carbon credits. %This analysis is necessary to implement a reliable trade market scheme in future versions of the cBSMD.

First of all, the average speed was calculated through travel distance and speed. Although this approach is not as accurate as the use of telemetry to get the driving behavior related GHG emission data, it is sufficient to get the commensurability of users. 
Based on the Ontario's vehicle model distribution, the \emph{carbon dioxide equivalent} (\ce{CO2e}) factor/mile of the models, the average speed, and the travel distance, the trip could be assigned to a GHG emission amount. In addition, the GHG emissions of bus trips were calculated with the \emph{Toronto Transit Commission} (TTC) bus fleet distribution and their associated CO2e factor/mile. %Note that more detailed models that consider drive cycles can enhance the accuracy of the estimated GHG emissions  \cite{mcnerney2017}, yet the method above is sufficient for the cBSMD demonstration purpose here at stake.

In order to turn the GHG emission amount per trip into the GHG consumption per user, the GHG emission data was divided by number of passengers and average bus seats respectively. The TTC bus fleet distribution led to an average bus seat number of 50.55/bus. Furthermore, the amount of GHG emission per user was multiplied by the current \ce{CO2e} price, 20 CAD per tonne (in Ontario, Canada) as well as with a factor of $100$ to increase the scale of the final token and to avoid too many decimals respectively. As a result, 100 carbon tokens equal 1ct (CAD).

The total amount of carbon tokens on this particular day is $1,573,708.73$ (around $157.37$ CAD), distributed over $3,187$ users. Consequently, without considering any reduced cap, every user is entitled to consume $493.79$ tokens. The cBSMD simulates the credit's usage per user and time. Thus, it can displayed through the user ID and the assigned sociodemographic data which user has tokens left at the end of the day and could therefore sell there surplus to the other peers in the system who ran out of tokens. Moreover, it is possible to demonstrate trip patterns due to the cBSMD simulation.  

\subsection{Results}

In terms of computational resources consumed in the simulation, there are about $6$ token transactions per minute whereby the average token transaction has a latency of $0.037$sec with a standard deviation of $0.023$sec. A $100\%$ throughput was reached during the whole simulation. In \cite{Farooq2019} it is shown that using small power devices in a physical network, the BSMD (in which the cBSMD is build upon) can handle $4036$ transaction per minute with an average latency of $5.5$sec (standard deviation of $1.73$sec) and $99\%$ throughput.

%The results of the case study are divided into two sections; (1) the sociodemographic analysis (see Figure~\ref{fig:user_charts}) and (2) the trip-based analysis (see Figure~\ref{fig:trip_charts}). 

\begin{figure}
    \centering
    \includegraphics[width=0.80\textwidth]{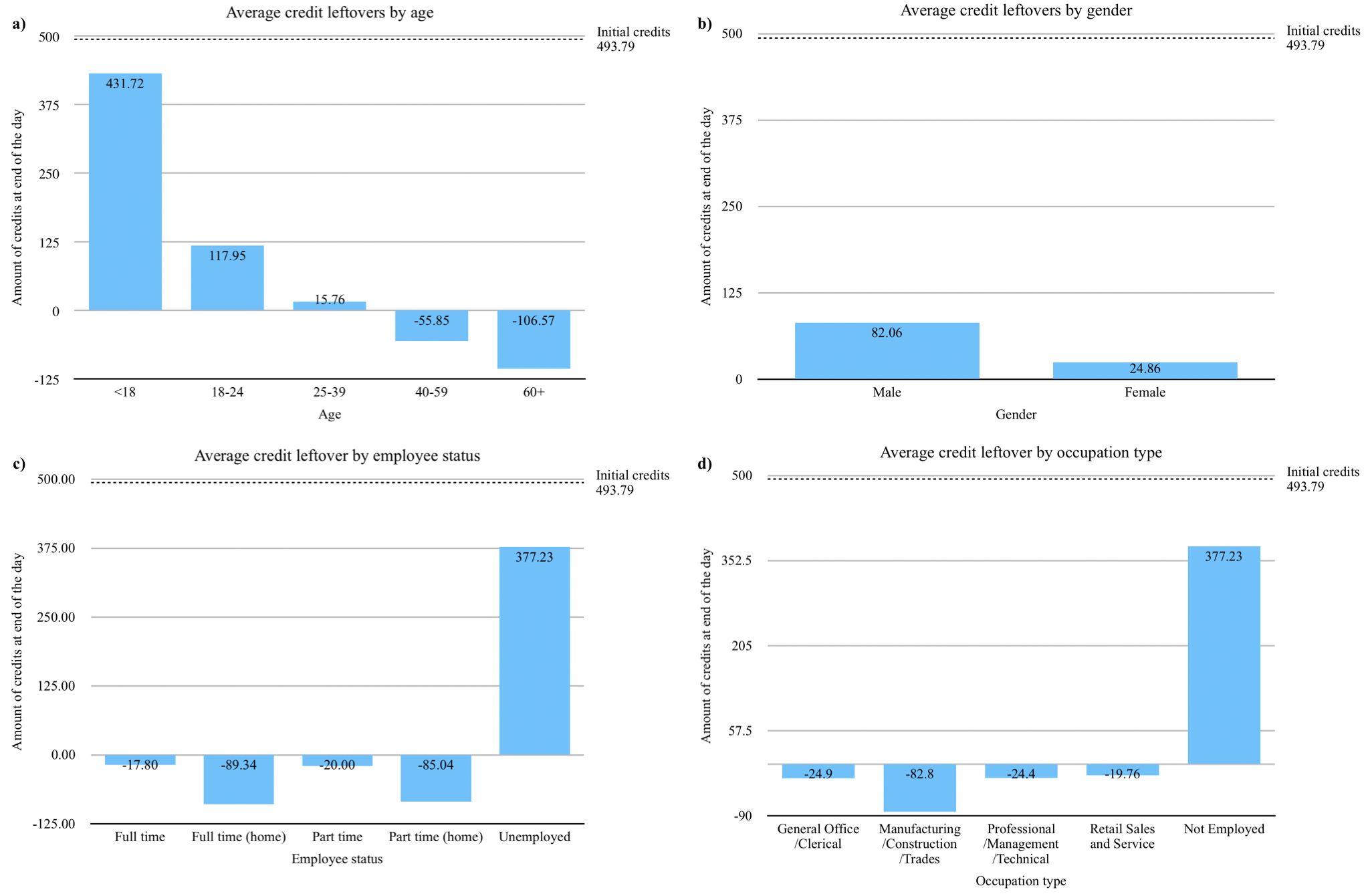}
    \includegraphics[width=0.80\textwidth]{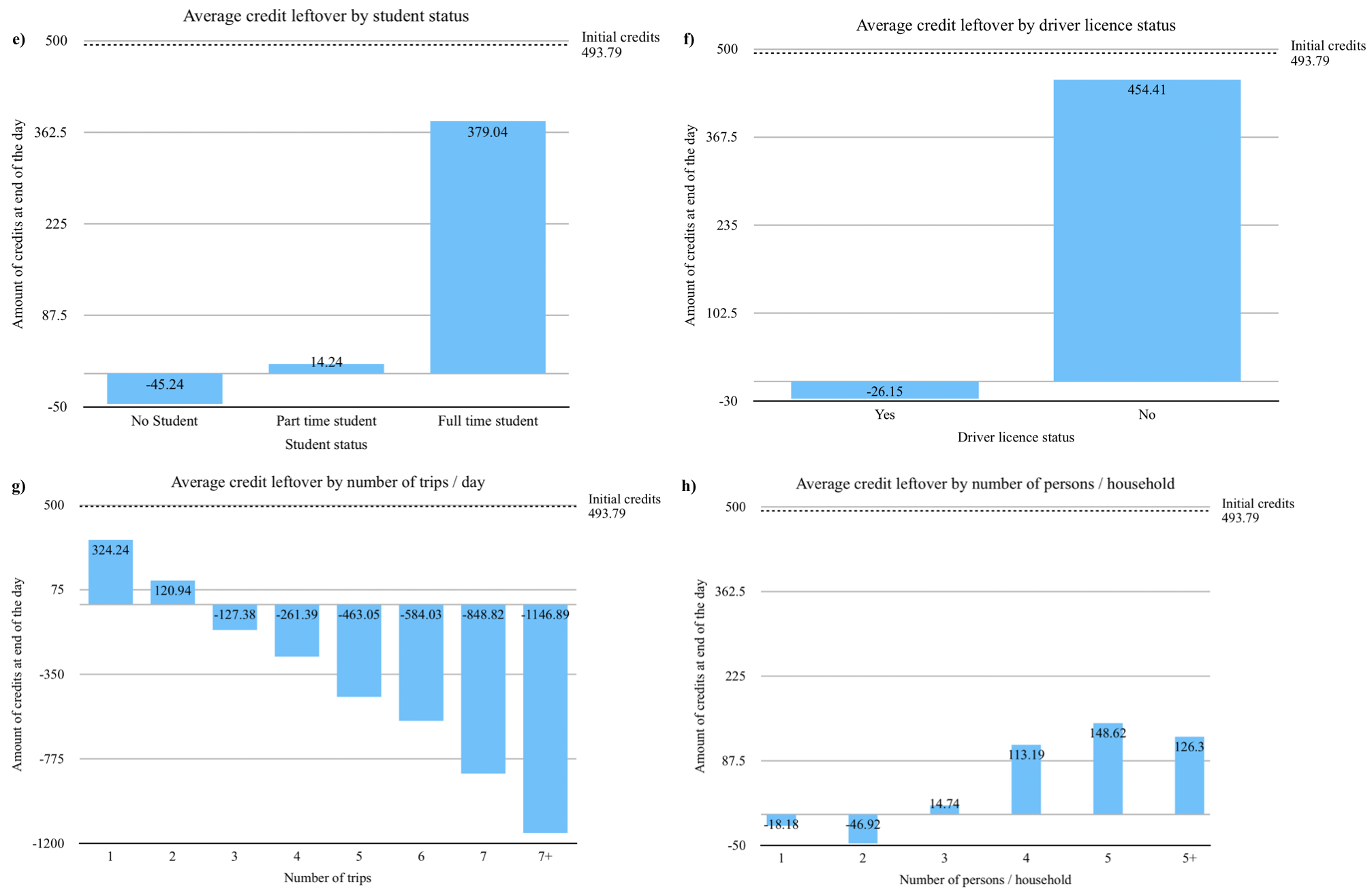}
    \includegraphics[width=0.80\textwidth]{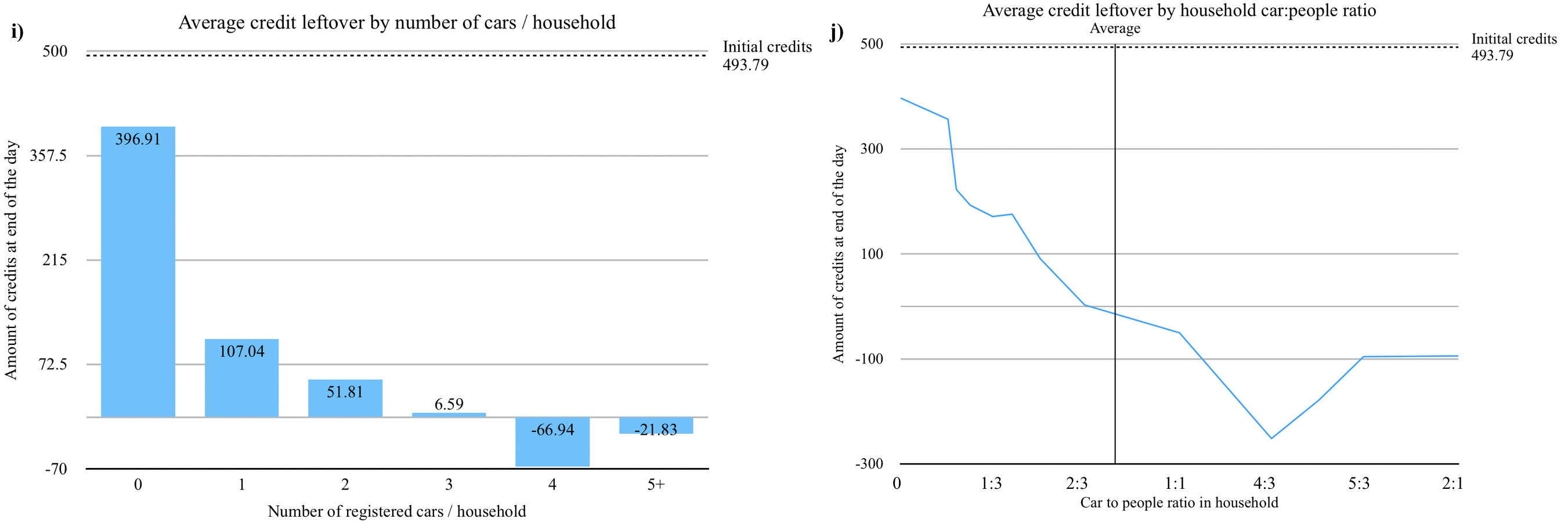}
    \caption{Sociodemographic analysis of token usage in the cBSMD}
    \label{fig:user_charts}
\end{figure}

In Figure~\ref{fig:user_charts} chart (a) is illustrated the distribution of the average token leftovers by age. It can be seen that the token usage increases steadily with growing age. This is caused by mainly two factors: 
First, the variety of the used modes. While persons under 18 use primarily the school bus ($39.84\%$ and walks ($39.01\%$) for their trips, the percentage of trips by private vehicles are increasing accordingly the age. Hence $98.39\%$ of persons with an age of $60$ and above use the private car for their trips. The growth of the usage of the car does increase enormously between the group of $18-24$ years old persons ($69\%$) and the $25-39$ years old persons ($94.05\%$). 

Second, the number of passengers on car trips. While users with an age of $60$ and above have a car passenger average of $1.21$ on their car trips, the groups with also a high share of car trips do have a higher average of passengers, $1.49$ ($25-39$ years old persons) and $1.42$ ($40-59$ years old persons). This is caused by the fact that the younger the users are the more mandatory trips they have to do as going to, for instance, school, university, work. Therefore, the user have to switch to alternative modes when they do not have access to a private car. This is also crucial for the outcome of a higher number of average passengers. Users share more rides when they know what they have to do during the day (e.g. share the ride to the workplace, bring the children to school/afternoon activity), this pattern is for the most retired persons not the case. 

In Figure~\ref{fig:user_charts}(b), the token usage by gender is revealed. The males have in average $82.06$ tokens left at the end of day, the females $24.86$. In order to analyze the causes for this pattern, the trip distances and trip amounts of the day have to be evaluated. Females do travel more often than males, $2.46$ to $2.36$, and travel longer distances, $13,853$ to $12,618$. In addition, males do travel in total $3.73\%$ more emission free than females ($1.66\%$ more walks and $2.07\%$ more cycling trips). Thus, the higher average GHG emissions by women can be explained through the additional activities during the day. Around $2/3$ of the child care tasks are done by women in Canada. In addition, purchases and services are conducted in a distribution of $59:41$ (mothers to fathers) \cite{Houle2017}.

In Figure~\ref{fig:user_charts}(c) are shown the distribution of carbon tokens based on the employee status. Hereby, it is highlighted that unemployed users have the least share of GHG emissions and therefore the highest surplus of tokens at the end of the day. This is also caused by the fact that students and persons under the age of 11 are considered as unemployed. However, it is worth to mention that individuals who work from home, whether or not full time or part time do consume more tokens during the day than users who have to commute to their workplace. 
When a user works from home, the trip amount during the day is higher than the number of trips of commuters. In fact, employees at home: full time ($3.11$ trips, $14,\-248$ distance), part time ($3$ trips, $14,\-466$) and employees who have to commute: full time ($2.39$ trips, $15,\-943$ distance), part time ($2.73$ trips, $15,\-394$). In last, the unemployed users with an average number of trips of $1.9$ trips and an average travel distance of $8,\-237$.

In Figure~\ref{fig:user_charts}(d), the occupation type analysis is demonstrated. The as unemployed considered users follow the same pattern like in Figure~\ref{fig:user_charts}(c). Nevertheless, persons who work in \emph{General Office/Clerical}, \emph{Professional/Management/Technical} or \emph{Retail Sales and Services} do share the approximately same amount of carbon credits. Employees at a \emph{Manufacturing/Construction/Trades} workplace consume around four times as much the other employed users due to a larger travel distance, in average $2$ km ($1.24$ miles), and a greater average speed of $3.62$ km/h ($2.25$ mph). This could be caused by the workplace's location. 

In Figure~\ref{fig:user_charts}(e), the amount of token leftovers regarding the student status of the user is stated. It is shown that full time students do consume a far smaller GHG amount than part time and non-students. This entails an average token leftover of $379.04$ (\emph{full time students}), $14.24$ (\emph{part time students}) and $-45.24$ (\emph{non-students}). The reason for that is the mode distribution of \emph{full time students} Therefore, the school bus ($31.54\%$) and walks ($33.93\%$) constitute around 2/3 of the mode choice by full time students. In contrast the groups of \emph{part time students} and \emph{non-students}, whereby over $90\%$ of the users consider the car for their trips.
Furthermore, \emph{full time students} travel in average significantly less kilometers than \emph{part time students} and \emph{non-students}. In particular, $8,\-127$ (\emph{full time students}) compared with $13,\-754$ (\emph{part time students}) and $14,\-796$ (\emph{non-students}).

Figure~\ref{fig:user_charts}(f) illustrates the impact of the driver licence status on the GHG emissions. With $454.41$ to $-26.15$ leftovers it is shown that the most people who are entitled to drive the car for their journey, do that. Users without a driver licence are forced to take alternative modes and can save in this way carbon tokens.

Figure~\ref{fig:user_charts}(g) illustrates the correlation between number of trips and GHG consumption. Hence the more trip a user takes during the day, the more tokens she has to use in average. Accordingly to the increasing number of trips, the travel distance raises as well. However, the travel distance is not the only decisive factor in terms of the surge of GHG emissions. It is conspicuous that the distribution of used modes decreases with the increasing number of trips. As a result, $72.30\%$ of the users who complete one trip during the day use the car, $14.68\%$ a walk and $6.32\%$ the bus. Out of all users who do two, $79.44\%$ are car users, $7.59\%$ take the school bus and $6.36\%$ a walk. When a user complete three or more trips the distribution decreases rapidly, $95.48\%$ the car, $1.81\%$ the bus and $1.20\%$ a walk. 

In Figure~\ref{fig:user_charts}(h) demonstrates the token leftover regarding the number of registered persons per household at the end of the day. As a result, a correlation between GHG emission and household size is highlighted. The bigger the household, the bigger the token leftover and consequently smaller the GHG emission share. 
Hereby, three indicators have to be stated. Although the average number of trips does not contain any considerable differences, the average traveled distance vary between the users of different household sizes. Therefore, the users in a single household travel on average $6,938$ meters. The individual user in a two person household $11,902$ meters and the users who are part of three or more person households travel on average $14,230$ meters. 
The second reason for the differences in token consumption is the amount of passengers on the car-based trips. The bigger the household, the bigger the average number of passengers in the car. In particular, (1) $1.16$; (2) $1.19$; (3) $1.33$; (4) $1.59$; (5) $1.59$; (6+) $1.65$. In consequence, the emitted GHG due to car trips can divided on average by more persons when the household is bigger. 
The third crucial factor for the reliance between size of household and average GHG consumption is the variation of used modes. Thus, the bigger the household the higher the distribution of used modes. Hence, the distribution from a single household user - car ($92.79\%$), school bus ($0\%$), walks ($1.8\%$), bus ($2.7\%$), ride-hailing service ($2.7\%$), bicycle ($0\%$) - varies from a user who is living in a, for example, five person household - car ($60.23\%$), school bus ($18.56\%$), walks ($13.64\%$), bus ($5.68\%$), ride-hailing service ($0\%$), bicycle ($1.89\%$) - and is much more lopsided. 

In Figure~\ref{fig:user_charts}(i) shows the average token leftovers by number of registered cars per household. As a result, we can ascertain that the greater the number of registered cars the higher the individual's GHG consumption. This can be ascribed to the fact that users of households with more registered vehicles do not have to share the car as often as users with less cars per household. Consequently, the average number of passengers decreases by the surge of registered cars. 
Moreover, the percentage of car uses increases and the variation of modes decreases due to a higher number of registered cars on the household of the user. Users without a registered car on their household have a mode distribution of car ($9.09\%$), school bus ($0\%$), walks ($36.36\%$), bus ($27.27\%$), ride-hailing service ($18.18\%$), bicycle ($9.09\%$). 
Users with one registered car on their household: car ($71.84\%$), school bus ($5.83\%$), walks ($11.41\%$), bus ($8.25\%$), ride-hailing service ($0.24\%$), bicycle ($2.43\%$). 
Users without two registered car on their household: car ($76.92\%$), school bus ($9.41\%$), walks ($8.3\%$), bus ($3.87\%$), ride-hailing service ($0.08\%$), bicycle ($1.42\%$). 
Users with three registered car on their household: car ($84.26\%$), school bus ($3.79\%$), walks ($8.16\%$), bus ($1.75\%$), ride-hailing service ($0.29\%$), bicycle ($1.75\%$). And so one, whereby the percentage of users who take the car for their trips increases steadily regarding a higher number of registered cars of the users household.  

Figure~\ref{fig:user_charts}(j) highlights the average token leftovers by the ratio of registered cars to number of persons in a household. It is illustrated that the smaller the ratio, the more GHG emissions are produced per user. This is caused by the fact that the users do have more often the option to take the car instead of an emission-less alternative mode as well as the amount of passengers per car trip decreases accordingly. As a result, the users with a ration of $1$ car to $6$ persons have an average passenger amount of $2,16$, users in a household with a ratio of $1:1$ have $1.14$ passengers on average in the car.

\begin{figure}
    \centering
    \includegraphics[width=1\textwidth]{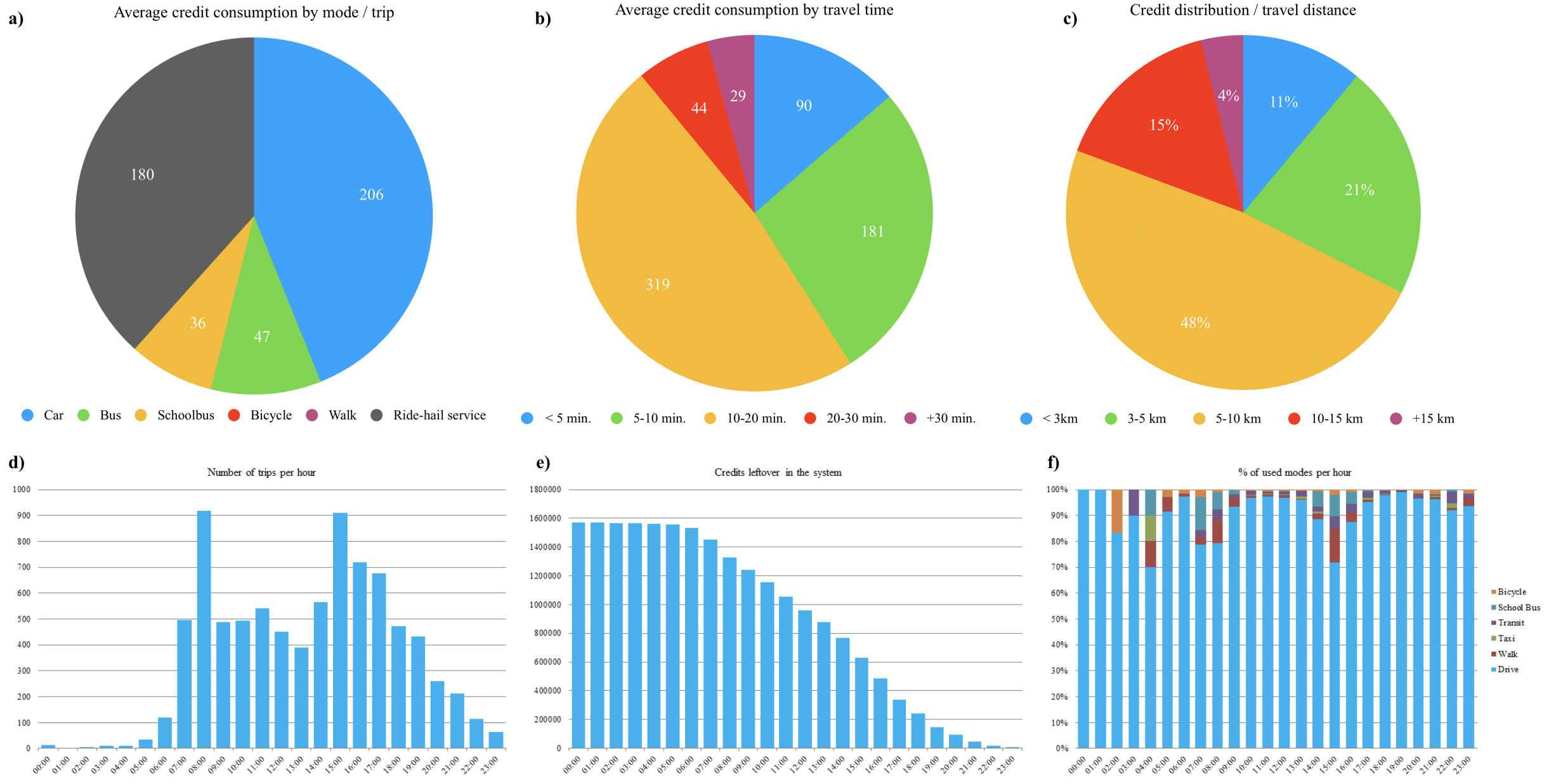}
    \caption{Trip-based analysis of token usage in the cBSMD}
    \label{fig:trip_charts}
\end{figure}

Figure~\ref{fig:trip_charts}(a) illustrates the fact that users produce the most GHG emissions due to the use of cars. The passengers in cars as well as in ride-hail services consume on average $206$ and $180$ carbon tokens, respectively. The public bus users, with 47 tokens, and the school bus users, with $36$ tokens, produce, based on the per seat charging system, only a fraction of active and passive car users. 

Figure~\ref{fig:trip_charts}(b) highlights the token consumption by travel time. Most users consume their tokens for travels between 10 and 20 minutes ($319$), independently of the choice of mode. Individuals who travel between $5$ and $10$ minutes do consume the double amount ($181$) of users who travel less than 5 minutes ($90$). Surprisingly, the amount of needed tokens for trips between $20$ and $30$ as well as longer than $30$ minutes is relatively low. This is caused through the use of different modes and their resulting lower average speed, for instance, bus or bicycle. 

Figure~\ref{fig:trip_charts}(c) illustrates the distribution of token usage by travel distance. As a consequence, it is shown that half of the allocated tokens were used by the users for trips between $5$ and $10$km ($3.1$ and $6.2$ miles). Furthermore, around a quarter of all tokens were used for trips between 3 and 5km ($1.9$ and $3.1$ miles).

Figure~\ref{fig:trip_charts}(d) describes the number of trips per hour. It is highlighted that between $8$ and $9$ a.m. as well as between $3$ and $4$ p.m. the number of trips is the greatest. This is attributed to the daily commuters between home and workplace. 

Figure~\ref{fig:trip_charts}(e) demonstrates the consumption of carbon tokens in the system by hour. Hence the total amount of token leftovers can be seen. A significant surge of consumed tokens is illustrated between $6$ a.m. and $10$ p.m.

In Figure~\ref{fig:trip_charts}(f) the percentage mode's variety is presented. Thus, within the commuting times, in morning and afternoon, the users tend to spread over more different types of mode.

\section{Conclusion and future work}
The \emph{carbon Blockchain framework for Smart Mobility Data-market} (cBSMD) is designed to provide individuals the accurate GHG footprint from a multi-modal perspective, which is necessary to implement a user-based GHG trading scheme. The cBSMD uses the \emph{Blockchain framework for Smart Mobility Data-market} (BSMD) as a basis to develop a user-based, tokenized GHG trading scheme. In addition to the existing BSMD, the proposed conceptual framework adds trip caused GHG emissions information to the trip data of the user. Consequently, a monitoring, reporting and verification of the emitted pollution can be ensured and utilized for the ETS. 
We demonstrate the usage of the cBSMD due to a simulation on a physical network with a blockchain composed of $3186$ \emph{passive} nodes that represent traveling users during the day as well as $4$ \emph{active} nodes which are responsible for hosting and running the cBSMD. All users in the system are allocated with the same average amount of tokens to cover their total emitted GHG. The allocated tokens are reduced based on emitted GHG whenever a user completes their trip. The case study provides an overview about the distribution of tokens during the day and did not include any ETS market setting and rules, individual behaviour when dealing with such a market, nor alternative token (travel) charging scheme. Based on the sociodemographic data, patterns could be established and the production of GHG emissions and consumption of tokens analyzed with cBSMD. Thus, the analysis gives us insights into who consumed carbon tokens, when, where and for what mode or mobility service.
To be able to use the system in real-life cases, similar sociodemographic analysis can be used to ensure the operationalizability of different token allocation and trading schemes. Consequently, a market-ready user-centric ETS under cBSMD should be tested in different experiment and simulation studies for behaviour, market, emission and scalability evaluation. This cBSMD demonstration showed its capability of materializing ETS strategies such that not only industries are forced to reduce their \ce{CO2e} emissions, but also travellers should be stimulated to manage their own GHG footprint in their mobility patterns. 

\section{Authors contribution statement}

Bilal Farooq and Carlos Lima Azevedo supervised the design and implementation of this study, provided input with regards to the conceptual framework, interpretations of results, and provided revision to the manuscript. David L\'opez contributed in the development of the case study and the token transactions over the blockchain. Johannes Eckert implemented the presented idea, developed the conceptual framework, performed the computations for the case-study and created the analysis of results. 

\newpage

\bibliographystyle{unsrt}
\bibliography{trb_template}
\end{document}